\documentclass[10pt,aps,prb,onecolumn]{revtex4-2}
\usepackage{graphicx,latexsym}
\usepackage{dcolumn}
\usepackage{amssymb,amsmath,bm}
\usepackage{color}
\newcommand{\beq}{\begin{equation}}
\newcommand{\eeq}{\end{equation}}

\begin{document}

\title{Generalized Jones Calculus for Vortex, Vector, and Vortex-Vector Beam Transformations}

\author{Wen-Hsuan Kuan}
\email{wenhsuan.kuan@gmail.com}
\author{Kuei-Huei Lin}
\author{Chang-Wen Wang}
\author{Ni-Hsuan Hung}
\affiliation{Department of Applied Physics and Chemistry, University of Taipei, Taipei, Taiwan}

\date{\today}

\begin{abstract}
The work defines the general form of Jones vector and establishes the Jones matrix for polarizers, wave plates, Faraday rotators, Q-plates, and spiral phase plates. We establish the generalized Jones calculus for vortex, vector, and vortex-vector beam transformations. Systematic formalism presents the theoretical arrangements in the manipulation of the phase and polarization of the structured lights. The eigenstates and the time-reversal behaviors of optical components and systems are discussed through matrix algebra. The generalized Jones calculus is then used in the characterization of Sagnac interferometers, and the potential applications are proposed.
\end{abstract}

\maketitle

\section{Introduction}
Electromagnetic waves characterized by their frequency, amplitude, phase, and polarization can be derived from Maxwell equations with given boundary and initial conditions. 
Recently, {optical vortex beams} with non-planar helical wavefronts are most attractive due to their unique properties and applications {\cite{Yao2011, Wang2016, Zhu2019}}.
The vortex beam has {spatially inhomogeneous phase} with the phase singularity along the beam axis, which reveal that in addition to spin angular momentum, the photons can also carry non-zero orbital angular momentum (OAM).
Recently, the studies on electric field distributions with {inhomogeneous polarization} are also blooming, and many unconventional optical phenomena have been reported {\cite{Wang2020}}. The typical example of the inhomogeneous polarization distribution is the {cylindrical vector beams} (CVB), such as radially- or azimuthally-polarized Laguerre-Gaussian (LG) modes.)
The {CVBs} are the paraxial solutions of vectorial Helmholtz equations, which have indeterminate state of polarization at the propagation axis; that is, there exists a singularity. When a spiral phase is added to the CVB, it acquires nonzero OAM and is termed as a {vortex vector beam}.

%
However, the theoretical modeling of singular optics is usually mathematically complicated.
Due to the strong spatial inhomogeneity close to the polarization singularity and the phase singularity, the commonly used dipole-moment approximation would be no longer valid for the description of the optical responses exerted by an optical vortex. These singularities have reinforced the re-examination of the gauge transformation and gauge invariance in the foundations of electromagnetic theories {\cite{Quinteiro2014}}. If the handedness of circular polarization and the OAM of the vector beam are not parallel, the magnetic parts of the light beam become of significant importance, and an interaction Hamiltonian which only accounts for electric fields is inappropriate {\cite{Quinteiro2014, Quinteiro2017}}. Similarly, the studies of the interaction of an LG beam with {multipolar atomic transitions} revealed the profound contributions of the electric and magnetic dipole vectors and the gradient of high-order electromagnetic fields in the determination of spatial-dependent transition rates {\cite{Klimov2012}}.
Combining the characteristic focusing property and the discrepancy in the relaxation time scale between conduction electrons and localized magnetic moments, the utilization of cylindrical vector beams is capable of developing the nonequilibrium analog of the magnetic oscillation measurement, which is difficult with the implementation of Gaussian beams {\cite{Fujita2018}}.

The establishment of the matrix calculus may avoid the integral equation or scattering theory in dealing with the light-matter interactions.
For example, the propagation of a Gaussian beam can be uniquely described via the successive application of the ABCD transformation. By applying the Collins-Huygens integral, the propagation of a doughnut beam generated by a spiral phase plate can also be analytically described
by an ABCD lens system {\cite{Mawardi2011}}.

%
Polarization is vital for how light behaves in photonic crystals, metals, metamaterials and how light is guided and distributed.
In terms of ellipse orientation and ellipticity, the application of four-parameter {Stokes vectors} can geometrically represent the state of polarization on the Poincar\'{e} sphere.
The formalism of the Stokes parameters can be used to describe the superposition of several polarized beams, provided that the beams are incoherent to each other.
Instead, Muller matrix formalism is appropriate when concerning the interaction of the polarized beams with polarizing devices.
It is also convenient to apply the {wave-transfer matrix and scattering matrix methods} to discuss the light transmission, reflection, and refraction through the dielectric multi-layers {\cite{Rumala2015}}.

On the other hand, when there are amplitude and phase relations between light beams, the utilization of the two-component Jones vectors established on the horizontal and vertical polarization bases will be an appropriate alternative in describing the polarization state and phase of monochromatic plane waves. In this manner, the polarization state's change when passing through some optical devices, such as polarizers, retarders, and rotators, can be analyzed by Jones matrix formalism. The Jones calculus may concisely interpret the beam's conversion and propagation for beams with spatially homogeneous polarization. However, conventional formalism fails to describe optical fields having polarization singularities.

Recently, Matsuo established the matrix calculus for cylindrical vector beams and deduced the rotation formulas for angularly variant Jones vectors and matrices on polar coordinates {\cite{Matsuo2011}}. By using Mach–Zehnder arrangement, Alonso et al. realized the generation and control of angular momentum from non-uniformly polarized beams {\cite{Alonso2012}}. We recognize that the conventional coordinate transformation has to be modified when the fast or slow axis of the waveplate, such as the Q-plate, is associated with the azimuthal angle. Moreover, when the Jones matrix is used to deal with uni-directional propagation, the transformation formulas involving non-reciprocal elements may also fail. Besides, in contrast to determining the state of polarization simply by the relative amplitude and phase of two polarization components where the Jones vectors with different weights are equivalent, the effect of the relative optical path length to the matrix formalism concerning beam superposition will be prominent.

In this paper, we propose establishing a generalized Jones calculus for vortex and vector and beam optics, in which the bidirectional wave propagation and superposition will be discussed within the Sagnac interferometer. Conventionally, non-adjustable optical components, such as a linear polarizer, half-wave plate, Q-plate, spiral-phase plate, etc., are used in the experimental setups for vortex beam or vector beam generation. However, it is preferred that the polarization state or optical angular momentum can be adjusted in some circumstances. This might be achieved by translating or rotating some of the components, but it suffers from the optical system's alignment and stability, especially for the free-space setups. Magneto-optical devices can modify the polarization state of light via an externally applied magnetic field without modifying the arrangement of the optical elements. Moreover, the change of polarization is associated with the propagation direction of light. Therefore, magneto-optical devices have significant potential for the generation and manipulation of the vortex and vector beams. As a result, taking advantage of the noncommutative relation of optical elements and superposition of beams may provide more flexibility or reduce the complexity in the design of vector and vortex beam relative optical systems.

\section{Formulation}
Assume that the oscillation phase of the electric field is $\omega t - kz + \delta$, and the azimuthal coordinate is $\phi$. The Jones vectors and matrices are represented with Cartesian basis, where the electric field components are referenced to the $x$ and $y$ axes. While focusing on the polarization and phase behaviors of electromagnetic waves, a structured beam with arbitrary polarization and phase can be generally described by
\begin{equation}
\mathbf{J}(\phi) \equiv\left(\begin{array}{l}
u(\phi) e^{i F(\phi)} \\
v(\phi) e^{i G(\phi)}
\end{array}\right),
\end{equation}
in which $u(\phi)$ and $v(\phi)$ denote the angular dependent polarization component and $e^{i F(\phi)}$ and $e^{i G(\phi)}$ depict the complex phase functions. The rotation of Jones vectors by a definite angle $\theta$ should follow
\begin{equation}
\mathbf{J}^{\prime}(\phi)=\mathbf{R}(\theta) \mathbf{J}(\phi-\theta)=\left(\begin{array}{cc}
\cos \theta & -\sin \theta \\
\sin \theta & \cos \theta
\end{array}\right) \mathbf{J}(\phi-\theta).
\end{equation}
On the other hand, as Jones matrix of an optical device can be generally written in the form of
\begin{equation}
\mathbf{M}(\phi) \equiv\left(\begin{array}{ll}
A(\phi) & B(\phi) \\
C(\phi) & D(\phi)
\end{array}\right),
\end{equation}
the expression $J_f(\phi) = M(\phi) J_i(\phi)$ has to be reformulated when the operation is performed by a rotated optical device whose fast axes are   azimuthally dependent, such as the Q-plate. Assuming that $J_i^r(\phi,\theta) = R(\theta) J_i(\phi)$ and $J_f^r(\phi,\theta) = R(\theta) J_f(\phi)$ interpret the state vectors under the rotation of coordinates, then we must arrive at
\begin{equation}
J_f^r(\phi,\theta) = R(\theta)M(\phi)R^{-1}(\theta)J_i^r(\phi,\theta)\equiv M'(\phi,\theta)J_i^r(\phi,\theta),
\end{equation}
in which
\begin{equation}
\mathbf{M}^{\prime}(\phi,\theta) = \left(\begin{array}{cc}
\cos(\phi + 2\theta) & \sin(\phi + 2\theta) \\
\sin(\phi + 2\theta) & -\cos(\phi + 2\theta)
\end{array}\right).
\end{equation}
However, it would be necessary to apply the change of variable $\phi \rightarrow \phi+\theta \equiv \phi'$ to preserve the azimuthal symmetry under rotation. As a consequence, the rotation of Jones matrix by a definite angle $\theta$ should be formulated following
\begin{eqnarray}
\mathbf{M}^{\prime}(\phi,\theta) &=& \mathbf{R}(\theta) \mathbf{M}(\phi-\theta) \mathbf{R}(-\theta) \\
&=& \left(\begin{array}{cc}
\cos(\phi + \theta) & \sin(\phi + \theta) \\
\sin(\phi + \theta) & -\cos(\phi + \theta)
\end{array}\right).
\end{eqnarray}
In this manner, an incident scalar beam with horizontal polarization parallel to the fast axis of Q-plate will transform to a vector beam with radial polarization, which will gradually vary with the rotation of the Q-plate and to the azimuthally polarized state as the rotation angle reaches $\theta = \pi/2$. We summarize some common Jones vectors and matrices in Table~\ref{table-jones}.
%
%
%
%
\begin{table}[t!]
\small
\caption{This is a table}
\label{table-jones}
\begin{tabular}{p{3cm}c p{4cm} c}
\hline
& Jones vector &  & Jones matrix \\
\hline
Horizontal polarization & $\left(\begin{array}{l} 1 \\
0\end{array}\right)$ & Linear polarization with Horizontal transmission axis & $\left(\begin{array}{ll}1 & 0 \\
0 & 0\end{array}\right)$ \\
Vertical polarization & $\left(\begin{array}{l}0 \\
1\end{array}\right)$ & Linear polarization with Vertical transmission axis & $\left(\begin{array}{ll}0 & 0 \\
0 & 1\end{array}\right)$ \\
Right (Left)-Handed Circular polarization & $\left(\begin{array}{l}1 \\
\mp i\end{array}\right)$ & Linear polarization with transmission axis tilted by $\alpha$  & $\left(\begin{array}{cc}
\cos^2\alpha & \sin\alpha\cos\alpha \\
\sin\alpha\cos\alpha & \sin^2\alpha
\end{array}\right)$ \\
Radial polarization & $\left(\begin{array}{c}
\cos \phi \\
\sin \phi
\end{array}\right)$ & Quarter-wave plate with Horizontal fast-axis & $\left(\begin{array}{cc}
1 & 0 \\
0 & i
\end{array}\right)$ \\
Azimuthal polarization & $\left(\begin{array}{c}
-\sin \phi \\
\cos \phi
\end{array}\right)$  & Half-wave plate with Horizontal fast-axis & $\left(\begin{array}{cc}
1 & 0 \\
0 & -1
\end{array}\right)$ \\
Horizontal polarization with optical angular momentum $\pm l$ & $\left(\begin{array}{c}
e^{\pm i l \phi} \\
0
\end{array}\right)$  & Q-plate with fast axis tilted by $\phi/2$  & $\left(\begin{array}{cc}
\cos \phi & \sin \phi \\
\sin \phi & -\cos \phi
\end{array}\right)$ \\
Vertical polarization with optical angular momentum $\pm l$ & $\left(\begin{array}{c}
0 \\
e^{\pm i l \phi}
\end{array}\right)$  & Faraday rotator &
$\left(\begin{array}{cc}
\cos \phi_0 & -\sin \phi_0 \\
\sin \phi_0 & \cos \phi_0
\end{array}\right)$ \\
& & Spiral phase plate with right (left)-handed azimuthal thickness increase & $\left(\begin{array}{cc}
e^{\pm i l \phi} & 0 \\
0 & e^{\pm i l \phi}
\end{array}\right)$ \\
\hline
\end{tabular}
\end{table}

\section{Special Wave-Plate and Beam Transformation}
%
While $M^{-1}_{Q} = M^T_{Q} = M^*_{Q}$, the Jones matrix of the Q-plate is orthogonal, unitary, and normal, which is also real symmetric since $M_{Q}M^T_{Q} = I$. Evaluating the eigenvalues of $M_{Q}$ gives $\varepsilon = \pm 1$, leading to the corresponding eigenvectors written by
\begin{equation}\label{eqn-Qeigen}
\left(\begin{array}{c}
\cos\phi/2\\
\sin\phi/2
\end{array}\right) \quad \textrm{or} \quad
\left(\begin{array}{r}
-\sin\phi/2\\
\cos\phi/2
\end{array}\right),
\end{equation}
respectively. The vectors in Eq.~(\ref{eqn-Qeigen}) reveal the azimuthally-dependent half-angle and vector beam converter characteristics of the Q-plate. However, the determinant $|M_{Q}| = -1$ indicates that its linear map is a combination of reflection and rotation, and would reverse the orientation. Therefore, $M_{Q}$ preserves no symmetry under time reversal and the polarization state of the beam passing backward through a Q-plate is neither preserved.

%
Optical vortices are characterized by phase dislocations or singularities due to the continuous spatial nature of the field.
The most direct way to create a helical wavefront is to allow the light beam to propagate into a medium with spiral inhomogeneity in the longitudinal direction to generate an integer phase step along the azimuthal angle {\cite{Beijersbergen1994}}. A simple way is to fabricate a plate with a helical surface called a spiral phase plate (SPP). It was demonstrated experimentally that a spiral phase plate could convert a TEM$_{00}$ laser beam into a helical-wavefront beam with a phase singularity at its axis {\cite{Sanchez2009}}. This optical element can change the phase distribution of a mode and benefit from the fact that it is polarization insensitive. Therefore, the matrix presentation of the SPP operator can be reduced from a tensor simply to a scalar. However, since $M^{-1}_{SPP} \neq M^T_{SPP} \neq M^*_{SPP}$, $M_{SPP}$ is not orthogonal, unitary, nor normal. Neither the operator preserves time-reversal invariant.
In addition to the matrix algebra, Bovino introduced a quantum-like representation of a spiral phase plate, which acts on an electromagnetic field as a two-mode phase operator {\cite{Bovino2011}} proving an alternative analysis of the optical operators.

%

The materials act as polarization rotators in the presence of an external magnetic field $\bm{B}$ is known as the Faraday magneto-optical effect {\cite{Haider2017, Lu2014}}. The angle of rotation $\phi_0 = VBL$ is proportional to the material's thickness $L$, and the rotatory power $\rho = V B$ is proportional to the magnetic density flux in the direction of wave propagation, where $V$ is called the Verdet constant. The Jones matrix of a Faraday rotator can be phenomenologically represented by
\begin{equation}
M_{F}(\phi_0) = \left(\begin{array}{cc}\label{eq-Faraday}
\cos\phi_0 & -\sin\phi_0 \\
\sin\phi_0 & \cos\phi_0
\end{array}\right).
\end{equation}
Equation (\ref{eq-Faraday}) indicates that the eigenstates of a polarized medium are circularly polarized, which can also be derived by solving the material equation from which the right (left)-circularly polarized beams propagating with a different effective refractive index would be observed.
Unlike the optical activity effect, the Faraday effect does not depend on the wave vector $\bm{k}$, so that reverse in the direction of propagation does not reverse the sense of rotation of the plane of polarization.

%
%
The transformation between a vortex beam and a vector beam can be achieved by sequentially employing a SPP and a Q-plate, i.e.,
\begin{equation}
R(\theta)M_{Q}(\phi-\theta)R(-\theta)\,M_{L-SPP}(\phi)
\left(\begin{array}{c}
e^{il\phi}\\
0
\end{array}\right) = \left(\begin{array}{l}
\cos(\phi+\theta)\\
\sin(\phi+\theta)
\end{array}\right).
\end{equation}
We would obtain radially polarized state $(\cos\phi\quad \sin\phi)^T$ by choosing $\theta = 0$ or azimuthally polarized state $(-\sin\phi \quad \cos\phi)^T$ by tuning $\theta = \pi/2$. Further, applying the relation $(\hat{A}\hat{B})^\dag = \hat{B}^\dag \hat{A}^\dag$, the vortex beam can be retrieved from a vector beam by the inverse transformation process:
\begin{equation}
M_{R-SPP}(\phi)\,R(-\theta)M_{Q}(-\phi+\theta)R(-\theta)
\left(\begin{array}{l}
\cos\phi\\
\sin\phi
\end{array}\right) = \left(\begin{array}{c}
e^{il\phi}\\
0
\end{array}\right).
\end{equation}
%

%

%
%
Upon a plane transverse to the propagation direction, we can define the orientation of the electric field and a reference axis $L_0$ of its phase. For linearly scalar beam or linearly polarized plane wave, $L_0$ is irrelevant to the location $\varsigma:=(x,y)$ on the transverse plane and neither temporal relevant. On the other hand, by utilizing wave plates and the beam splitter, we can draw the reference axes $L_1[+t]$ and $L_1[-t]$ rotating clockwise or counterclockwise to time, respectively. Further, we can obtain spatially- and temporally-dependent reference axes $L_2[{+\varsigma,+t}]$ and $L_2[{-\varsigma,-t}]$ utilizing inhomogeneous medium, such as Q-plate or spiral phase plate. In particular, we find that the temporal dependence of the reference axes can be ingeniously eliminated from the superposition resultant of two outgoing counter beams. In this manner, the spatial dependence of the reference axes is converted to that of the electric field's orientation or phase, leading to the generation of a fixed electric field reference axis $L_3$ and the vector beam, vortex beam, or vortex-vector beam.


\begin{figure}[t!]
\begin{center}
\includegraphics[scale=0.2]{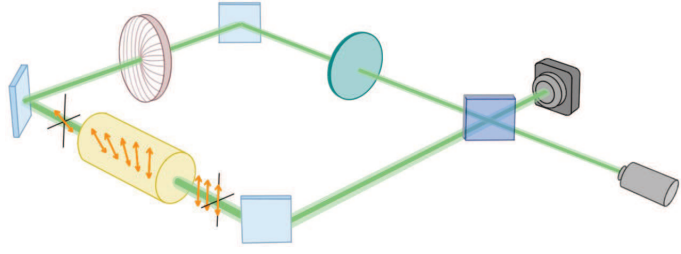}
\caption{Schematics of the Sagnac interferometer consisting of magneto-optical device, half-wave plate, and Q-plate.}
\label{fig:schematic-sagnac-2}
\end{center}
\end{figure}

\begin{figure}[b!]
\begin{center}
\includegraphics[scale=0.35]{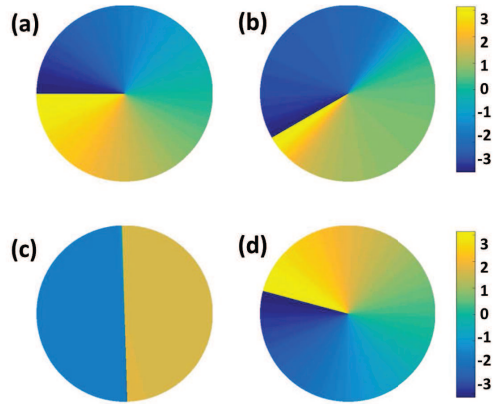}
\caption{The phase portraits of optical vortices with topological charge evolving from (a) $l = -\hbar$ to (d) $l = \hbar$. In (b) and (c), the emergence of the non-periodical structure due to the destructive interference of counterpropagating beams smashes the quantum signature of vorticity.}
\label{fig:vortex}
\end{center}
\end{figure}
%
\section{Sagnac Interferometer}
\subsection{Generation of Optical Vortices}

For an electric field initially prepared in the linearly polarized state, the generation of an optical vortex can be realized throughout a proper arrangement of the Sagnac interferometer shown in {Fig.~\ref{fig:schematic-sagnac-2}}.
Firstly, the beam is transformed into a circularly polarized state using a quarter wave-plate and then split into two orthogonal components: transmission and reflection via a beam splitter. The Jones vectors of these two components are given by

\begin{equation}
E_1^0 = \left(\begin{array}{l}
1 \\
i
\end{array}\right) \quad
E_2^0 = \left(\begin{array}{l}
-1 \\
-i
\end{array}\right).
\end{equation}
Circulating around a Sagnac interferometer consisting of magneto-optical device, half-wave plate, and Q-plate, the state of polarization of $E_1^0$ is modified to
\begin{eqnarray}
E_{1}^{out} &=& M_{\lambda/2}R(\theta)M_Q(\phi-\theta)R(-\theta)M_F(\phi_0)E_1^0 \nonumber\\
&=& \left(\begin{array}{l}
\cos(\phi-\phi_0+\theta) - i\sin(\phi-\phi_0+\theta) \\
i\cos(\phi-\phi_0+\theta) - \sin(\phi-\phi_0+\theta)
\end{array}\right).
\end{eqnarray}
On the other hand, the anti-circulation modifies the state of polarization of $E_2^0$ to
\begin{eqnarray}
E_{2}^{out} &=& M_{F}(-\phi_0)R(-\theta)M_Q(-\phi+\theta)R(\theta)M_{\lambda/2}E_2^0 \nonumber\\
&=& \left(\begin{array}{l}
-\cos(\phi+\phi_0+\theta) - i\sin(\phi+\phi_0+\theta) \\
-i\cos(\phi+\phi_0+\theta) + \sin(\phi+\phi_0+\theta)
\end{array}\right).
\end{eqnarray}
The superposition of $E^{out}_{1}$ and $E^{out}_{2}$ gives
\begin{equation}
E^{out}= \left(\begin{array}{l}
-2i \sin\phi_0\, e^{-i ( \phi + \theta )} \\
2 \sin\phi_0\, e^{-i ( \phi + \theta )}
\end{array}\right),
\end{equation}
which is a circularly polarized scalar beam carrying an integer OAM $l = -\hbar$ {and cane be filtered to a pure optical vortex with a linear polarizer.}

It is interesting that if we adjust the relative phase of the incident circularly polarized beam components and the elements' order and guide the resultant wave passing through a polarizer (axis finder) before it is imaged, we can achieve manipulating the sign of the topological charge by adjusting the magnetic field strength and rotation angles of the Q-plate and polarizer. For example, by choosing $\phi_0 = -\pi/4$, $\theta = \pi/3$, and $\alpha = \pi/2$, the Jones vector can be a vertically polarized field carrying a spiral phase $e^{i(\phi+\pi/12)}$, whereas the inverse magneto-optical effect can induce the handiness change of the spiral wave. The phase portraits of optical vortices with topological charge evolving from $l = -\hbar$ to $l = \hbar$ is shown in {Fig.~\ref{fig:vortex}}(a)-(d). In between, due to counterpropagating beams' destructive interference, the non-periodical phase patterns emerge and smash the quantum signature of vorticity.

\begin{figure}[b!]
\begin{center}
\includegraphics[scale=0.2]{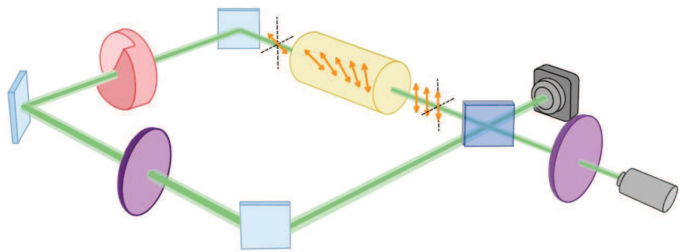}
\caption{Schematics of the Sagnac interferometer consisting of magneto-optical device, spiral phase plate, and quarter-wave plate.}
\label{fig:schematic-sagnac-1}
\end{center}
\end{figure}
%
\subsection{Generation of Vector Beams}
\begin{figure}[t!]
\begin{center}
\includegraphics[scale=0.35]{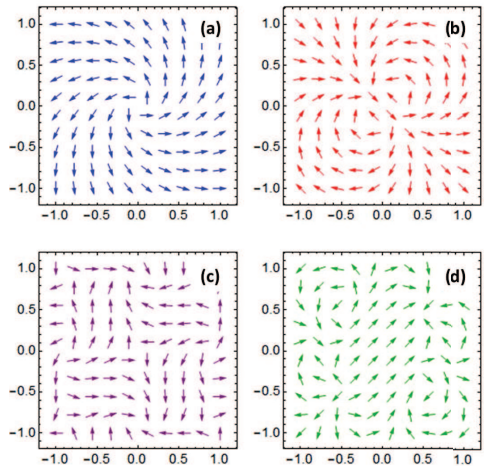}
\caption{Planar distribution of topologically-induced vector fields correspond to (a) $l = $, (b) $l = 2$, (c) $l = 3$, and (d) $l = 4$.}
\label{fig:vector}
\end{center}
\end{figure}

For an electric field initially prepared in the linearly polarized state, it is also possible to generate vector beams throughout a proper arrangement of the Sagnac interferometer shown in {Fig.~\ref{fig:schematic-sagnac-1}}.
For example, set the Jones vectors of the transmission and reflection components as
\begin{equation}
E_1^0 = \left(\begin{array}{l}
1 \\
-1
\end{array}\right) \quad
E_2^0 = \left(\begin{array}{l}
-1 \\
1
\end{array}\right).
\end{equation}
The modification of the state of polarization of $E_1^0$ circulating around a Sagnac interferometer consisting of magneto-optical device, spiral phase plate, and quarter-wave plate is given by
\begin{eqnarray}
E_{1}^{out} &=& M_{\lambda/4}M_{R-SPP}(\phi-\theta_{SPP})M_F(\phi_0)E_1^0 \nonumber\\
&=& \left(\begin{array}{l}
e^{il(\phi-\theta_{SPP})} [\cos(\phi_0) + \sin(\phi_0)] \\
i e^{il(\phi-\theta_{SPP})} [\cos(\phi_0) - \sin(\phi_0)]
\end{array}\right),
\end{eqnarray}
and after anti-circulating $E_2^0$ becomes
\begin{eqnarray}
E_{2}^{out} &=& M_F(-\phi_0)M_{\lambda/2}M_{L-SPP}(\phi-\theta_{SPP})M_{\lambda/4}E_2^0 \nonumber\\
&=& \left(\begin{array}{l}
-e^{-il(\phi-\theta_{SPP})} [\cos(\phi_0) + i\sin(\phi_0)] \\
e^{-il(\phi-\theta_{SPP})} [-i\cos(\phi_0) + \sin(\phi_0)]
\end{array}\right).
\end{eqnarray}
For $\phi_0 = \pi/2$ and $\theta_{SPP} = 0$, we obtain
\begin{equation}
E_{out}= (1 - i) \left(\begin{array}{l}
\cos l\phi - \sin l\phi \\
\cos l\phi + \sin l\phi
\end{array}\right),
\end{equation}
In {Fig.~\ref{fig:vector}} we demonstrate the planar distribution of vector fields for $l = \{1,4\}$. Unlike the optical vortices that characterize the angular phase variation, the polarization configuration is illustrated by mimicking spin textures domain walls as topological defects in the two-dimensional XY model {\cite{Toulouse1976}}. As a consequence, we can identify (a) to spiral around a vortex core (b) circulating two vortices, one with an up-oriented core and the other with a down-oriented core, and as chiral favored,  (c) establishing mirror-symmetric flows by repelling from a centered antivortex, and (d) spiraling off two vortices towards the sink established by the vortex-antivortex pairs.

\section{Conclusion}

In this work, we defined the general form of the Jones vector and established the Jones matrix for some optical elements. We established the generalized Jones calculus for vortex, vector, and vortex-vector beam transformations.
By employing the Sagnac interferometer consisting of special wave-plates and magneto-optical Faraday rotator, we have investigated the effects of the non-reciprocal elements and the relative optical path length to the matrix formalism and beam superposition in the bi-directional propagation. We have also analyzed the eigenstates and the time-reversal behaviors of optical components and systems through matrix algebra for the structured beam generation.
With a proper arrangement of magneto-optical device, half-wave
plate, and Q-plate, we realized the scalar to vortex beam transformation. We have also realized manipulating the topological charge sign by adjusting the magnetic field strength and rotation angles of the Q-plate and linear polarizer. We found that the counterpropagating beams' destructive interference would smash the quantum signature of vorticity. By tuning the rotation angle of the magneto-optical device, spiral phase plate, and quarter-wave plate consisting of the Sagnac interferometer, we have realized the synthesis of topologically-induced vector beams. Unlike the optical vortices that characterize the angular phase variation, the polarization configuration was identified by mimicking spin textures
domain walls as topological defects in the two-dimensional XY model.
Our study shows that taking advantage of the noncommutative relation of optical elements and superposition of beams may provide more flexibility or reduce the complexity in the design of vector and vortex beam relative optical systems.


\section*{Acknowledgements}
The authors thank partially financial support from the Ministry of Science and Technology (MOST), Taiwan (MOST 108-2221-E-845-004 and MOST 109-2221-E-845-002).


\begin{thebibliography}{9}
%
\bibitem{Yao2011} A. M. Yao and M. J. Padgett, "Orbital angular momentum: origins, behavior and applications," Adv. Opt. Photon. 3, 161 (2011).
%
\bibitem{Wang2016} J. Wang, "Advances in communications using optical vortices," Photon. Res. 4, B14 (2016).
%
\bibitem{Zhu2019} L. Zhu and J. Wang, "A review of multiple optical vortices generation: methods and applications," Front. Optoelectron. 12, 52 (2019).
%
\bibitem{Wang2020} J. W. Wang, F. Castellucci, and S. Franke-Arnold, "Vectorial light–matter interaction: Exploring
spatially structured complex light fields," AVS Quantum Sci. 2, 031702 (2020).
%
\bibitem{Quinteiro2014} G. F. Quinteiro, D. E. Reiter, and T. Kuhn, "Formulation of the twisted-light–matter interaction at the phase singularity: the twisted-light gauge," Phys. Rev. A 91, 033808 (2014).
%
\bibitem{Quinteiro2017} G. F. Quinteiro, D. E. Reiter, and T. Kuhn, "Formulation of the twisted-light–matter interaction at the phase singularity: beams with strong magnetic fields," Phys. Rev. A 95, 012106 (2017).
%
\bibitem{Klimov2012} V. V. Klimov, D. Bloch, M. Ducloy, and J. R. Rios Leite, "Mapping of focused Laguerre-Gauss beams: The interplay between spin and orbital angular momentum and its dependence on detector characteristics," Phys. Rev. A 85, 053834 (2012).
%
\bibitem{Fujita2018} H. Fujita and M. Sato, "Nonequilibrium Magnetic Oscillation with Cylindrical Vector Beams," Sci. Rep. 8, 15738 (2018).
%
\bibitem{Mawardi2011} A. Mawardi, S. Hild, A. Widera, and D. Meschede, Opt. Express 19, 21205 (2011).
%
\bibitem{Rumala2015} Y. S. Rumala, "Wave transfer matrix for a spiral phase plate," Appl. Opt. 54, 4395 (2015).
%
\bibitem{Matsuo2011} S. Matsuo, "Matrix calculus for axially symmetric polarized beam," Opt. Express 19, 12815 (2011).
%
\bibitem{Alonso2012} M. Alonso, G. Piquero, and J. Serna, "Proposals for the generation of angular momentum from non-uniformly polarized beams," Opt. Commun. 285, 1631 (2012).
%
\bibitem{Beijersbergen1994} M. W. Beijersbergen, R. P. C. Coerwinkel, M. Kristensen, J. P. Woerdman, "Helical-wavefront laser beams produced with a spiral phase plate," Opt. Commun. 112, 321-327 (1994).
%
\bibitem{Sanchez2009}, V. Ram\'{i}rez-S\'{a}nchez, G. Piquero, and M. Santarsiero, "Generation and characterization of spirally polarized fields," J. Opt. A: Pure Appl. Opt. 11, 085708 (2009).
%
\bibitem{Bovino2011} F. A. Bovino, "Quantum like representation of a spiral phase plate," arXiv:1104.2201 (2011).
%
\bibitem{Haider2017} T. Haider, "A Review of Magneto-Optic Effects and Its Application," Int. J. Electromagn. Appl. 7, 17 (2017).
%
\bibitem{Lu2014} X. Lu and L. Chen, "Vortex generation and inhomogeneous Faraday rotation of a nonparaxial Gaussian beam in isotropic magneto-optic crystals," Opt. Lett. 39, 3728 (2014).
%
\bibitem{Toulouse1976} G. Toulouse, M. Kleman, J. De Phys. Lett. 37, L149 (1976).
%
%
%
\end{thebibliography}
\end{document}